\documentclass[conference,a4paper]{IEEEtran}

\usepackage[utf8]{inputenc}
\usepackage[T1]{fontenc}
\usepackage{microtype}
\usepackage{scrextend}
\usepackage{caption}
\usepackage{subcaption}
\usepackage{paralist}
\usepackage[nocompress]{cite}

\usepackage{xspace}
\usepackage{relsize} 

\usepackage[printonlyused,nolist]{acronym}

\usepackage{amsmath}
\usepackage{amssymb}
\usepackage{stackrel}

\usepackage{balance}
\usepackage{paralist}
\usepackage[lowtilde]{url}
\usepackage{enumitem}

\renewcommand{\subsection}[1]{\smallskip\noindent{\bf{#1.}}}

\renewcommand{\subsubsection}[1]{\smallskip\noindent{\em\bfseries #1.}}

\usepackage{longtable}
\usepackage{booktabs}
\usepackage{multirow}
\usepackage{rotating} 

\usepackage{tabulary}

\usepackage{dcolumn}
\newcolumntype{d}{D{.}{.}{2.1}}
\newcolumntype{e}{D{.}{.}{3.1}}
\newcolumntype{f}{D{.}{.}{4.1}}
\newcolumntype{g}{D{.}{.}{5.1}}
\newcolumntype{h}{D{.}{.}{2.2}}
\newcolumntype{i}{D{.}{.}{3.2}}
\newcolumntype{j}{D{.}{.}{4.2}}
\newcolumntype{k}{D{.}{.}{5.2}}
\newcolumntype{m}{D{.}{.}{1.2}}

\usepackage[table]{xcolor}
\usepackage{graphicx}
\usepackage{wrapfig} 
\graphicspath{{figures/}}

\usepackage{tikz}
\usetikzlibrary{arrows}
\usetikzlibrary{chains}
\usetikzlibrary{fadings}
\usetikzlibrary{fit}
\usetikzlibrary{positioning}
\usetikzlibrary{shapes}

\usepackage[noend]{algorithmic}
\usepackage{algorithm}

\makeatletter
\newcommand*\listingsize{\@setfontsize\listingsize{7.8}{9.4}}
\makeatother

\usepackage{fancyvrb}
\usepackage{listings}
\usepackage{fixltx2e}
\usepackage[rounded]{syntax}
\let\oldsdlengths\sdlengths
\renewcommand\sdlengths{\oldsdlengths\setlength{\sdfinalskip}{0pt}}

\usepackage[%
  pdfpagemode=UseNone,
  pdfpagelabels=false,plainpages=true,
  bookmarks=false,
  ]{hyperref}

\makeatletter
\renewcommand\fs@ruled{\def\@fs@cfont{\bfseries}\let\@fs@capt\floatc@ruled
  \def\@fs@pre{\kern4pt\hrule height.8pt depth0pt \kern2pt}%
  \def\@fs@post{\kern0pt\hrule\relax\vspace{-5mm}}%
  \def\@fs@mid{\kern2pt\hrule\kern2pt}%
  \let\@fs@iftopcapt\iftrue}
\makeatother

\newcommand\definetool[2] {\newcommand{#1}{{\smaller\sc #2}\xspace}}
\definetool{\blast}       {Blast}
\definetool{\cpachecker}  {CPAchecker}
\definetool{\cbmc}        {Cbmc}
\definetool{\cil}         {Cil}
\definetool{\llvm}        {LLVM}
\definetool{\tvla}        {Tvla}
\definetool{\ocaml}       {OCaml}
\definetool{\tvp}         {Tvp}
\definetool{\camplp}      {CamlP4}
\definetool{\foci}        {Foci}
\definetool{\tcp}         {TCP}
\definetool{\escjava}     {ESC/Java}
\definetool{\slam}        {Slam}
\definetool{\jpf}         {JPF}
\definetool{\sycmc}       {SyCMC}
\definetool{\impact}      {Impact}
\definetool{\wolverine}   {Wolverine}
\definetool{\ufo}         {Ufo}
\definetool{\mathsat}     {MathSAT}
\definetool{\smtinterpol} {SMTInterpol}

\newcommand{\CPA}{{CPA}\xspace}
\newcommand{\cegar}{{\ac{CEGAR}}\xspace}
\newcommand{\CFA}{{\ac{CFA}}\xspace}

\newcommand{\SMT}{{\ac{SMT}}\xspace}
\newcommand{\safe}{{\sc true}\xspace}
\newcommand{\unsafe}{{\sc false}\xspace}

\newcommand{\false}{\mathit{false}}
\newcommand{\seq}[1]{{\langle #1 \rangle}}
\newcommand{\sem}[1]{[\![ #1 ]\!]}

\newcommand{\locs}{\mathit{L}}
\newcommand{\op}{\mathit{op}}
\newcommand{\pc}{\mathit{l}}

\newcommand{\pci}{{\pc_0}}

\newcommand{\pct}{{\pc_{e}}}

\newcommand{\concr}{{\cal C}}
\newcommand{\defran}{\textrm{def}}
\newcommand{\varAssignment}{v}
\newcommand{\cpaSymbol}{\mathbb{D}}
\newcommand{\interpret}[2]{{#1_{/#2}}}
\newcommand{\eval}[2]{{#1_{/#2}}}
\renewcommand{\path}{\sigma}
\newcommand{\cseq}{{\gamma}}
\newcommand{\cseqintpol}{{\Gamma}}

\newcommand{\Ints}{\mathbb{Z}}

\newcommand{\transconc}[1]{\smash{\stackrel{#1}{\rightarrow}}}

\newcommand{\wait}{\mathsf{waitlist}}
\newcommand{\reached}{\mathsf{reached}}

\renewcommand{\implies}{\Rightarrow}

\newcommand{\precisions}{\Pi}
\newcommand{\programprecisions}{\locs \to 2^\precisions}
\newcommand{\pr}{\pi}

\newcommand{\pto}{\mbox{$\;\longrightarrow\!\!\!\!\!\!\!\!\circ\;\;\;$}}

\newcommand{\SP}[2]{{\mathsf{SP}_{#1}({#2})}}

\newcommand{\mytitle}{Domain-Type-Guided Refinement Selection\\ Based on Sliced Path Prefixes}
\newcommand{\myauthor}{Dirk Beyer, Stefan Löwe, and Philipp Wendler}

\hypersetup{
	pdfauthor={\myauthor},
	pdftitle={\mytitle},
}

\begin{document}
\setcounter{page}{0}

\thispagestyle{empty}
\begin{minipage}{17cm}
\begin{center}
~\\[3cm]
\Huge{\mytitle}
\\[2cm]
\large{\myauthor}
\\[1cm]
\normalsize
{University of Passau, Germany}\\[7cm]

\hspace{-5mm}
\includegraphics[scale=0.2]{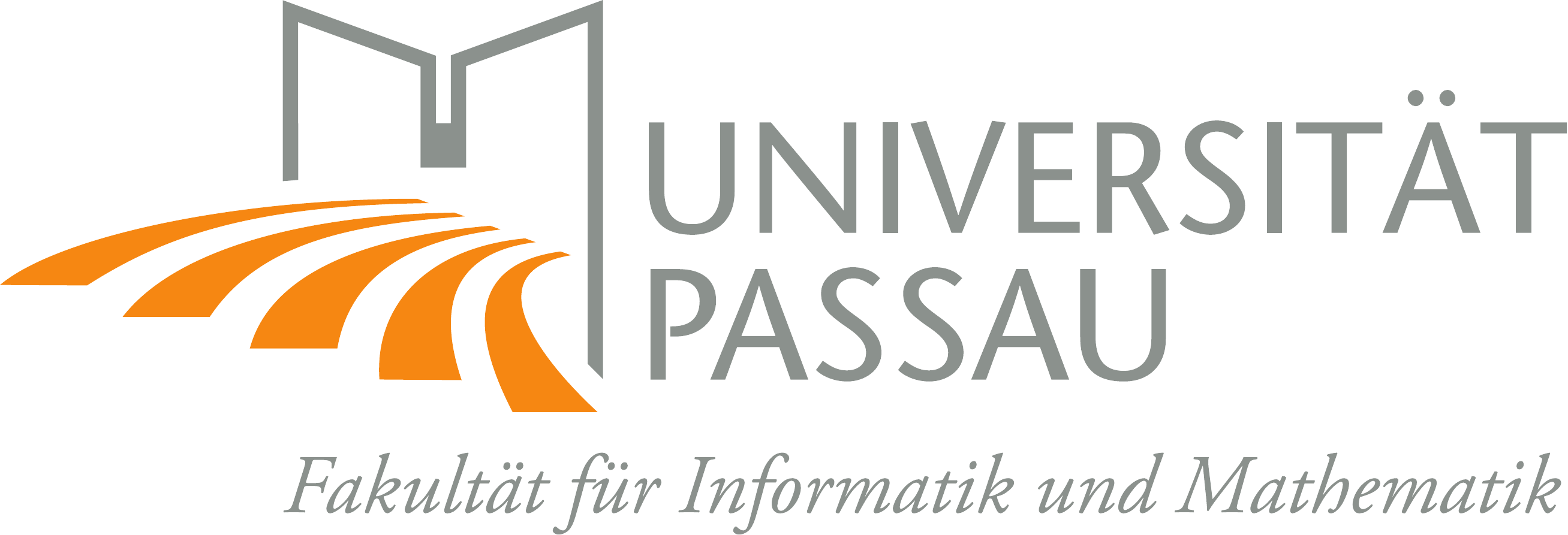} \\[1cm]
Technical Report, Number MIP-1501\\
Department of Computer Science and Mathematics\\
University of Passau, Germany\\
January 2015
\end{center}
\end{minipage}

\title{\mytitle}

\author{{\myauthor}
  \vspace{3mm}\\
  {University of Passau, Germany}\\
}
\maketitle
\thispagestyle{plain}
\pagestyle{plain}

\begin{abstract}
  %
Abstraction is a successful technique in software verification,
and interpolation on infeasible error paths is a successful approach to
automatically detect the right level of abstraction in
counterexample-guided abstraction refinement.
%
Because the interpolants have a significant influence
on the quality of the abstraction,
and thus, the effectiveness of the verification,
an algorithm for deriving the best possible interpolants
is desirable.
%
We present an analysis-independent technique that makes it possible to extract several alternative sequences of interpolants
from one given infeasible error path, if there are several reasons for infeasibility in the error path.
We take as input the given infeasible error path and apply a slicing technique to obtain
a set of error paths that are more abstract than the original error path but still infeasible, each for a different reason. 
The (more abstract) constraints of the new paths can be passed to a standard interpolation engine,
in order to obtain a set of interpolant sequences, one for each new path.
The analysis can then choose
from this set of interpolant sequences and select the most appropriate,
instead of being bound to the single interpolant sequence
that the interpolation engine would normally return.
For example, we can select based on domain types of variables in the interpolants,
prefer to avoid loop counters, or compare with templates for potential loop invariants,
and thus control what kind of information occurs in the abstraction of the program.
We implemented the new algorithm in the open-source verification framework \cpachecker
and show that our proof-technique-independent approach
yields a significant improvement of the effectiveness and efficiency of the verification process.
\acresetall

\end{abstract}

\section{Introduction}
In the field of automatic software verification,
abstraction is a well-understood and widely-used technique,
enabling the successful verification of real-world, industrial programs~(cf.~\cite{SLAMtransfer,LDV12,ASTREE}).
Abstraction makes it possible to omit certain aspects of the concrete semantics
that are not necessary to prove or disprove the program's correctness.
This may lead to a massive reduction of a program's state space,
such that verification becomes feasible within reasonable time and resource limits.
For example, \slam~\cite{SLAM} uses predicate abstraction~\cite{GrafSaidi97}
for creating an abstract model of the software.
%
One of the current research directions is to invent techniques to
automatically find suitable abstractions.
An ideal model is abstract enough to avoid state-space explosion
and still contains enough detail to verify the property.
\begin{figure}[b]
  \centering
    \includegraphics[width=\linewidth]{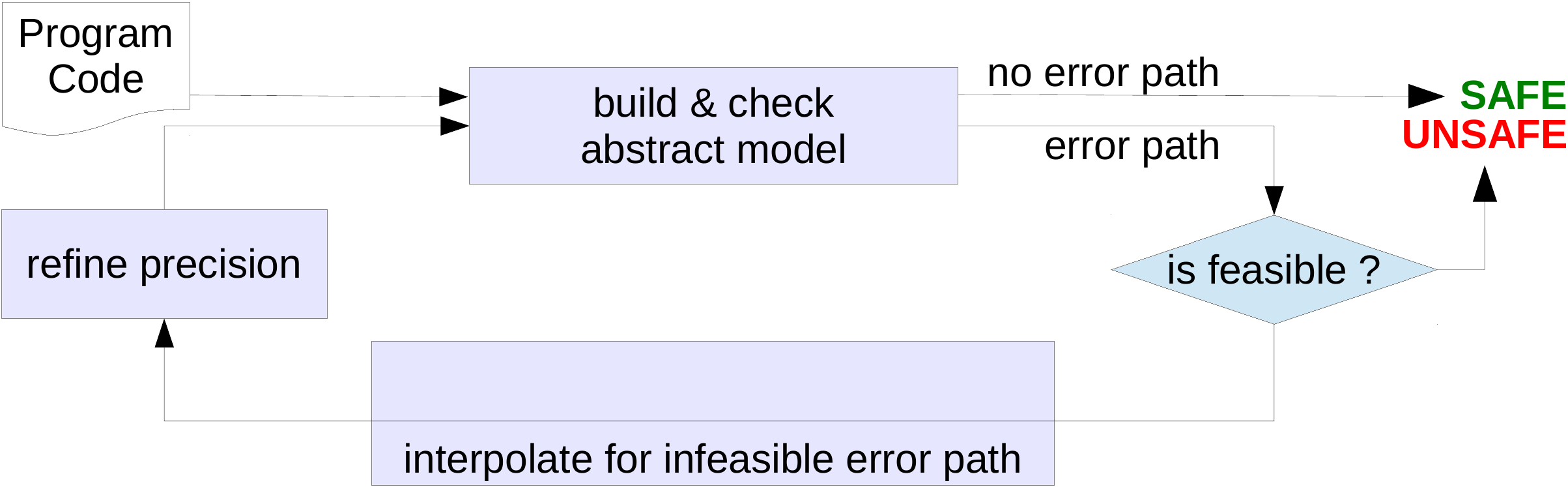}
    \caption{Example of the CEGAR loop, using a single error path for interpolation}
    \label{figure:singlePathItp}
    \vspace{-1mm}
\end{figure}
%
Counterexample-guided abstraction refinement (CEGAR)~\cite{ClarkeCEGAR} 
is an automatic technique that
starts with a very coarse abstraction and
iteratively refines an abstract model using \emph{infeasible} error paths
(witnesses of property violations).
If the analysis does not find an error path in the abstract model,
the analysis terminates and reports the verdict \safe
(property holds).
Because the abstract model over-approximates the program,
the verdict applies for the actual program.
If the analysis finds an error path, the path is checked for feasibility.
If the found error path does not contain a contradiction
and the error is indeed reachable
according to the concrete program semantics,
then the error path is feasible and a real error was found.
The analysis terminates and reports the verdict \unsafe
(program violates property).
If, however, the error path is actually infeasible,
then a ``spurious counterexample'' was found,
and the property violation is due to a too coarse abstract model.
The (contradicting) constraints of the infeasible error path 
can then be passed to an interpolation engine,
and the obtained interpolants identify information that is needed for refining the current abstraction,
such that the same infeasible error path is excluded in subsequent CEGAR iterations.
After refinement, the analysis proceeds with rebuilding a refined abstract model
in the next CEGAR iteration.
Several successful software verifiers
(e.g., \slam~\cite{SLAM}, \blast~\cite{BLAST}, \cpachecker~\cite{CPACHECKER}, \ufo~\cite{UFO})
make use of the CEGAR loop, which is illustrated in Figure~\ref{figure:singlePathItp}.

Craig interpolation~\cite{Craig57}
is a technique
that yields for two contradicting formulas an interpolant formula
that contains less information than the first formula,
but is still expressive enough to contradict the second formula.
This can be extended to a sequence of formulas.
In software verification,
interpolation was first used for the domain of predicate abstraction~\cite{AbstractionsFromProofs},
and later for value-analysis domains~\cite{CPAexplicit}.
Independent of the analysis domain,
interpolants for path constraints of infeasible error paths can be
used to refine abstract models and to eliminate the infeasible error paths 
in subsequent CEGAR iterations.
\begin{figure}[t!]
\begin{minipage}{.20\textwidth}
\begin{lstlisting}
extern int f(int x);
int main() {
  int b = 0;
  int i = 0;
  while(1) {
    if(i > 9) break;
    f(i++);
  }
  if(b != 0) {
    if(i != 10) {
      assert(0);
    }
  }
}
\end{lstlisting}
\end{minipage}
\hfil
\begin{minipage}{.05\textwidth}
  \includegraphics[scale=0.4]{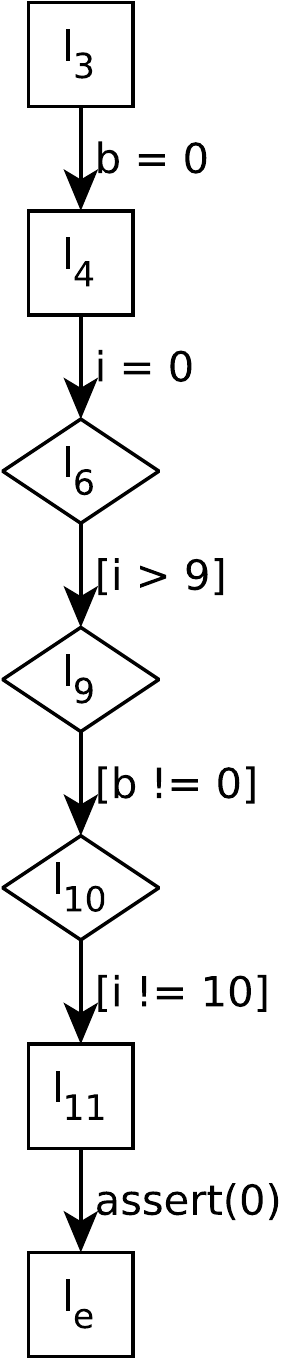}
\end{minipage}
\hfil
\begin{minipage}{.2\textwidth}
  \includegraphics[width=\linewidth]{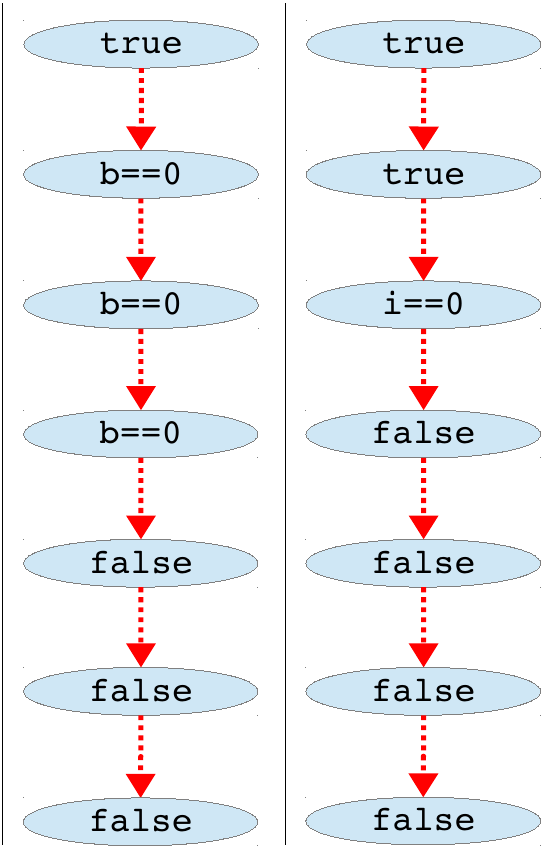}
\end{minipage}
\caption{Input program, with infeasible error path, and a ``good'' and a ``bad'' interpolant sequence}
\label{figure:goodBadItps}
\end{figure}
In this context, it is important to point out
that the choice of interpolants is crucial
for the performance of the analysis.
Figure~\ref{figure:goodBadItps} gives an example:
In this program, the analysis
will typically find the shown error path,
which is infeasible for two different reasons:
both the value of~\texttt{i}
and the value of~\texttt{b}
can be used to find a contradiction.
In general, it is now beneficial for the verifier
to track the value of the boolean variable~\texttt{b},
and not to track the value of the loop-counter variable~\texttt{i},
because the latter has many more possible values,
and tracking it would usually lead to an expensive unrolling of the loop.
Instead, if only variable~\texttt{b} is tracked,
the verifier can conclude the safety of the program without unrolling the loop.
Thus, we would like to get from the interpolation engine
the left shown interpolant sequence (only with boolean variable)
and not the right interpolant sequence (with loop-counter).
%
However, interpolation engines typically do not allow
to guide the interpolation process
towards a ``good'', or away from a ``bad'', interpolant sequence.
The interpolation engines inherently cannot do a better job here:
they do not have access to information
such as whether a specific variable is a loop counter and should be avoided in the interpolant.
Instead, which interpolant is returned
depends solely on the internal algorithms
of the interpolation engine.
This is especially true if the model checker in use
does not provide its own implementation of an interpolation engine
but rather makes calls to a library, e.g., a \SMT solver,
which normally cannot be controlled on such a fine-grained level.
In this case, the model checker is stuck
to what the interpolation engine returns,
be it good or bad for the verification process.
\begin{figure}[t!]
  \centering
  \includegraphics[width=\linewidth]{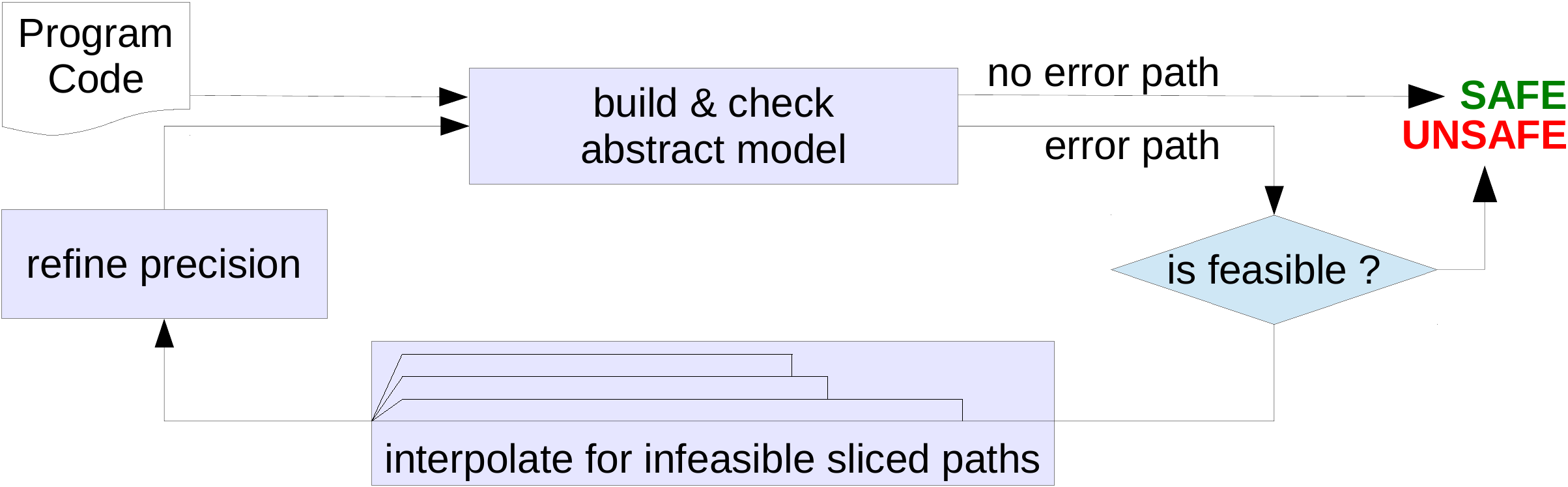}
  \caption{Example of the CEGAR loop, using a set of paths for interpolation}
  \label{figure:multiplePathsItp}
  \vspace{-2mm}
\end{figure}

Therefore, we present an approach
to \emph{guide} the interpolation engine to produce interpolants that we would like to get,
without changing the interpolation engine.
For this, we extract from an infeasible error path
a set of infeasible sliced paths stemming all from the same infeasible error path.
Each of these sliced paths can be used for interpolation,
yielding different interpolant sequences that are all expressive enough
to eliminate the original infeasible error path.
As depicted in Figure~\ref{figure:multiplePathsItp},
our approach fits well into CEGAR,
because only the refinement component needs customization,
and the new approach remains compatible with off-the-shelf interpolation engines.

\subsection{Contributions}
This paper makes the following key contributions:
\begin{itemize}
  \item we introduce a domain- and analysis-independent method to extract infeasible \emph{sliced} paths 
        from infeasible error paths,
  \item we prove that interpolants for such a sliced path are also interpolants for 
        the original infeasible error path,
  \item we explain that ---and how--- it is possible to obtain better interpolants 
        (in comparison to the standard approach) from a set of infeasible sliced paths,
        and that refinement selection plays a significant role in CEGAR,   
  \item we implement the presented concepts in the open-source 
        framework for software verification \cpachecker, and
  \item we show by experimental results that the novel approach to obtain better interpolants
        significantly improves the verification effectiveness and efficiency.
\end{itemize}

\subsection{Related Work}
The desire to control what interpolants an interpolation engine produces,
and trying to make the verification process more efficient
by finding good interpolants, is not new.
Our goal is to contribute a technique that
is independent from the abstract domain that a program analysis uses,
and independent from specific properties of interpolation engines.

The first work in this direction was suggesting to
make the interpolation configurable such that the user has a choice
between strong and weak interpolants, by controlling the
\emph{interpolant strength}~\cite{InterpolantStrength}.
This approach is unfortunately not implemented in the standard interpolation engines;
it requires to rewrite the algorithm that extracts interpolants from resolution proofs.
The technique of interpolation abstractions~\cite{ExploringInterpolants},
a generalization of term abstraction~\cite{LazyAbstractionForArrays},
can be used to guide solvers to pick good interpolants.
This is achieved by extending the concrete interpolation problem
by so called templates
(e.g., terms, formulas, uninterpreted functions with free variables)
to obtain a more abstract interpolation problem.
An interpolant for the abstract interpolation problem
is also a solution to the concrete interpolation problem.
Because these interpolation abstractions form a lattice,
suitable interpolants can be chosen
using a cost function.
Our approach is independent from the abstract domain
and interpolation engine, and does not rely on SMT solving.
For example, our technique is applicable to 
value and octagon domains.

Path slicing~\cite{PathSlicing05} is a technique that was introduced to reduce the burden
of the interpolation engine:
Before the constraints of the path are given to the interpolation engine,
the constraints are weakened by removing facts that are not important
for the infeasibility of the error path, i.e., a more abstract error path is constructed.
We also make the error path more abstract, but in different directions
to obtain different interpolant sequences, from which we can choose the ones that
yield the best abstract model.
While path slicing is interested in reducing the run time of the \emph{interpolation engine}
(by omitting some facts),
we are interested in reducing the run time of the \emph{verification engine}
(by spending more time on interpolation but creating a better abstract model).

\addtolength{\textfloatsep}{\baselineskip}
\vspace{\baselineskip}

\section{Background}
\label{section:background}
Our approach is based on several existing concepts,
and in this section we remind the reader of 
some basic definitions and our previous work in this field~\cite{CPAexplicit}.

\subsection{Programs, Control-Flow Automata, States, Paths, Precisions}
We restrict the presentation
to a simple imperative programming language,
where all operations are either assignments or assume operations,
and all variables range over integers\,%
\footnote{Our implementation is based on \cpachecker{},
which operates on C~programs;
non-recursive function calls are supported.}.
%
A program is represented by a \CFA.
A \CFA $A = (\locs, \pci, G)$ consists of a set~$\locs$ of program locations,
  which model the program counter,
an initial program location~$\pci \in \locs$,
  which models the program entry,
and a set $G \subseteq \locs \times Ops \times \locs$ of control-flow edges,
  which model the operations that are executed
when control flows from one program location to the next.
The set of program variables that occur in operations from~$Ops$
is denoted by~$X$.
%
A~\emph{verification problem}~${P = (A, \pct)}$ consists of a CFA $A$,
  representing the program,
and a target program location~$\pct \in \locs$,
  which represents the specification, i.e., ``the program must not reach location~$\pct$''.

A \emph{concrete data state} of a program is a variable assignment $cd: X \to \Ints$,
which assigns to each program variable an integer value;
the set of integer values is denoted as~$\Ints$.
A~\emph{concrete state} of a program is a pair~$(l, cd)$,
where $l \in \locs$ is a program location
and $cd$~is a concrete data state.
The set of all concrete states of a program is denoted by~$\concr$,
a subset~$r \subseteq \concr$ is called \emph{region}.
Each edge~$g \in G$ defines a labeled transition relation 
$\mathord{\transconc{g}} \subseteq \concr \times \{g\} \times \concr$.
The complete transition relation~$\transconc{}$ is the union
over all control-flow edges:
$\mathord{\transconc{}} = \bigcup_{g \in G} \transconc{g}$.
We write $c \transconc{g} c'$ if $(c, g, c') \in \mathord{\transconc{}}$,
and $c \transconc{} c'$ if there exists a $g$ with $c \transconc{g} c'$.

An \emph{abstract data state} represents a region
of concrete data states,
formally defined as abstract variable assignment.
An \emph{abstract variable assignment} is  
a partial function~${v: X \pto \Ints}$ or $\bot$,
where $v$~maps variables in its definition range
to integer values, and 
$\bot$ is used to represent no variable assignment
(i.e., no value is possible, similar to the predicate~$\false$ in logic).
The special abstract variable assignment~$\top = \{\}$ 
does not map any variable to a value and is used as initial abstract variable assignment
in a program analysis.
Variables that do not occur in the definition range of an
abstract variable assignment
are either omitted by purpose for abstraction in the analysis, or
the analysis is not able to determine a concrete value
(e.g., resulting from an uninitialized variable declaration or from an external function call).
For two partial functions~$f$ and $f'$, we write $f(x)=y$ for the predicate $(x, y) \in f$,
and $f(x)=f'(x)$ for the predicate $\exists c: (f(x) = c) \land (f'(x) = c)$.
We denote the \emph{definition range} for a partial function~$f$
as ${\defran(f) = \{ x \mid \exists y: f(x) = y\}}$,
and the \emph{restriction} of a partial function~$f$ 
to a new definition range~$Y$ as~$f_{|Y} = {f \cap (Y \times \Ints)}$.
%
An abstract variable assignment~$v$ represents the set~$\sem{v}$
of all concrete data states~$cd$ 
for which $\varAssignment$ is valid,
formally: ${\sem{\bot} = \{\}}$ and for all $v \not= \bot$,
$\sem{v} = {\{cd \mid \forall x \in X: 
v(x) = cd(x)\}}$.
%
The abstract variable assignment $\bot$ is called \emph{contradicting}.
The \emph{implication} for 
abstract variable assignments is defined as follows:
$v$~implies~$v'$ (written $v \implies v'$)
if $v = \bot$, or  
 for all variables $x \in \defran(v')$
 we have $v(x) = v'(x)$.
The \emph{conjunction} for
abstract variable assignments $v$ and~$v'$ is defined as:
\vspace{1mm}\\
$v \land v' =
\vspace{1mm}
\left\{
\begin{array}{ll}
    \bot      & \textrm{if } v = \bot \textrm{ or } v' = \bot\\ 
              &   \textrm{~~or } \exists x \in \defran(v) \cap \defran(v'): v(x) \not= v'(x)\\
    v \cup v' & \textrm{otherwise}
\end{array}
\right.$\\

The \emph{semantics of an operation} $\op \in Ops$ is defined
by the strongest-post operator~$\SP{\op}{\cdot}$:
given an abstract variable assignment~$\varAssignment$,
$\SP{\op}{\varAssignment}$ represents the set of concrete data states that are reachable
from the concrete data states in the set~$\sem{\varAssignment}$ by executing~$\op$.
%
Formally, given an abstract variable assignment~$v$
and an assignment operation~$x := exp$,
we have
$\SP{x := exp}{\varAssignment} = v_{|{X \setminus \{x\}}} \land v_x$ with
\vspace{1mm}\\
$v_x =
\vspace{1mm}
\left\{
\begin{array}{ll}
    \{ (x, c)\}  & \textrm{if } c \in \Ints \textrm{ is the result of the arithmetic } \\
                 & \textrm{~~evaluation of expression } \eval{exp}{v}\\
    \{\}         & \textrm{otherwise (if $\eval{exp}{v}$ cannot be evaluated)}
\end{array}
\right.$\\
where $\eval{exp}{v}$ denotes the interpretation of expression~$exp$ 
for the abstract variable assignment~$v$.
%
Given an abstract variable assignment~$v$
and an assume operation\,%
${[p]}$,
we have 
$\SP{[p]}{v} = \bot$ if $v = \bot$ or the predicate $\interpret{p}{v}$ is unsatisfiable,
or we have
\vspace{1mm}
$\SP{[p]}{v} = v \land v_p$ with
$v_p = \left\{(x, c) \in \left(X\setminus\defran(v) \times \Ints\right) ~\big|~ \interpret{p}{v} \implies (x = c)\right\}$,
\vspace{1mm}
and $\interpret{p}{v} = p \land 
  \bigwedge\limits_{y \in \defran(v)} y = v(y)$.

\enlargethispage{1mm}
A \emph{path}~$\path$ is a sequence~$\seq{(\op_1, \pc_1), \ldots, (\op_n, \pc_n)}$
of pairs of an operation and a location.
The path~$\path$ is called \emph{program path} if for every $i$ with ${1 \leq i \leq n}$
there exists a \CFA edge~${g = (\pc_{i-1}, \op_i, \pc_i)}$
and $\pc_0$ is the initial program location,
i.e., $\path$ represents a syntactic walk through the \CFA.
The result of appending the pair $(\op_n, \pc_n)$ to a path
$\path = \seq{(\op_1, \pc_1), \ldots, (\op_m, \pc_m)}$
is defined as $\path \land (\op_n, \pc_n) = \seq{(\op_1, \pc_1), \ldots, (\op_m, \pc_m), (\op_n, \pc_n)}$.
%
\linebreak
Every path~$\path = \seq{(\op_1, \pc_1), \ldots, (\op_n, \pc_n)}$ defines a \emph{constraint sequence}
$\cseq_\path = \seq{\op_1, \ldots, \op_n}$.
The \emph{conjunction}~${\cseq \land \cseq'}$ of two constraint sequences
${\cseq = \seq{\op_1, \ldots, \op_n}}$ and $\cseq' = \seq{\op_1', \ldots, \op_m'}$
is defined as their concatenation, 
i.e., ${\cseq \land \cseq' =  \seq{\op_1, \ldots, \op_n, \op_1', \ldots, \op_m'}}$,
the \emph{implication} of $\cseq$~and~$\cseq'$ (denoted by $\cseq \implies \cseq'$) 
as the implication of their strongest-post assignments
$\SP{\cseq}{\top} \implies \SP{\cseq'}{\top}$,
and $\cseq$~is \emph{contradicting} if $\SP{\cseq}{\top} = \bot$.
%
The \emph{semantics of a path}
$\path = \seq{(\op_1, \pc_1), \ldots, (\op_n, \pc_n)}$
is defined as the successive application of the strongest-post operator 
to each operation of the corresponding constraint sequence~$\cseq_\path$:
$\SP{\cseq_\path}{\varAssignment} = \SP{\op_n}{ \ldots \SP{\op_1}{\varAssignment} \ldots }$.
%
The set of concrete program states that result from running a program path~$\path$
is represented by the pair~$(l_n, \SP{\cseq_\path}{\top})$,
where $\top$ is the initial abstract variable assignment.
%
A~path $\path$ is \emph{feasible} if $\SP{\cseq_\path}{\top}$ 
is not contradicting, i.e., $\SP{\cseq_\path}{\top} \not= \bot$.
A~concrete state~$(l_n, cd_n)$ is \emph{reachable},
denoted by $(l_n, cd_n) \in Reach$,
if there exists a feasible program path $\path = \seq{(\op_1, \pc_1), \ldots, (\op_n, \pc_n)}$
with $cd_n \in \sem{\SP{\cseq_\path}{\top}}$.
A~location~$l$ is reachable if there exists a concrete data state~$cd$ such that $(l, cd)$~is reachable.
A~program is \emph{safe} (the specification is satisfied) if $\pct$ is not reachable.
A~path $\path = \seq{(\op_1, \pc_1), \ldots, (\op_n, \pct)}$, which
ends in $\pct$, is called \emph{error path}.

The \emph{precision} is a function $\pr: \programprecisions$,
where $\precisions$~depends on the abstract domain used by the analysis.
It assigns to each program location
some analysis-dependent information that defines the level of abstraction of the analysis.
For example, when using predicate abstraction,
the set~$\precisions$ is a set of predicates over program variables.
When using a value domain,
the set~$\precisions$ is the set~$X$ of program variables,
and a precision defines which program variables should be tracked by the analysis
at which program location.

\begin{algorithm}[t]
\vspace{1mm}
\begin{small}
\caption{\textbf{ }$\textsf{CEGAR}(\cpaSymbol, e_0, \pr_0)$, cf.~\cite{CPAexplicit}
         \label{algorithm:cegar}}
\begin{algorithmic} [1]
\INPUT
    a CPA with dynamic precision adjustment $\cpaSymbol$ and\\
    an initial abstract state~$e_0\in E$ with precision~$\pr_0\in(\programprecisions)$\\
\vspace{1mm}
\OUTPUT verification result \safe (property holds) or \unsafe
\vspace{1mm}
\VARDECL a set $\reached$ of elements of $E \times (\programprecisions)$,\\
    a set $\wait$ of elements of $E \times (\programprecisions)$, and\\
    an error path $\path = \seq{(\op_1, \pc_1), \ldots, (\op_{n}, \pc_{n})}$

\vspace{1mm}
\STATE $\reached := \{(e_0, \pr_0)\}$;
$\wait := \{(e_0, \pr_0)\}$;
$\pr := \pr_0$
\WHILE {$true$}
  \STATE $(\reached, \wait) := \textsf{\CPA}(\cpaSymbol, \reached, \wait)$
  \IF {$\wait = \{\}$}
    \STATE {\bf return } \safe
  \ELSE
    \STATE $\path := \mathsf{ExtractErrorPath}(\reached)$
    \IF [error path is feasible: report bug] {$\mathsf{IsFeasible}(\path)$}
      \STATE {\bf return } \unsafe
    \ELSE [error path is infeasible: refine and restart] 
      \STATE $\pr := \pr \cup \mathsf{Refine}(\path)$
      \STATE $\reached := \{(e_0, \pr)\}$;
             $\wait := \{(e_0, \pr)\}$
    \ENDIF
  \ENDIF 
\ENDWHILE
\end{algorithmic}
\end{small}
\vspace{1mm}
\end{algorithm}

\renewcommand{\cegar}{CEGAR\xspace}
\subsection{Counterexample-Guided Abstraction Refinement (CEGAR)}
\label{sec:cegar}
CEGAR
is a technique for automatic iterative refinement of an abstract model~\cite{ClarkeCEGAR}.
\cegar is based on three concepts:
(1) a \emph{precision}, which determines the current level of abstraction,
(2) a \emph{feasibility check}, deciding if an error path (the counterexample) is feasible, and
(3) a \emph{refinement} procedure, which takes as input an infeasible error path
    and extracts a precision to refine the abstract model such that
    the infeasible error path is eliminated from further exploration.
Algorithm~\ref{algorithm:cegar} shows an outline
of a generic and simple \cegar algorithm.
It uses the \textsf{\CPA} algorithm~\cite{CPAexplicit, CPAplus}
for program analysis with dynamic precision adjustment
and an abstract domain~$\cpaSymbol$ that is formalized as a
configurable program analysis (CPA) with dynamic precision adjustment.
The CPA uses a set~$E$ of abstract states and a set~$\programprecisions$ of precisions.
The analysis algorithm computes the sets $\reached$ and $\wait$,
which represent the current reachable abstract states with precisions
and the frontier, respectively.
The analysis algorithm is run first with $\pr_0$ as coarse initial precision
(usually $\pr_0(l) = \{\}$ for all $l \in L$).
If all program states have been exhaustively checked
and no error was reached,
indicated by an empty $\wait$,
then the \cegar algorithm terminates
and reports \safe (program is safe).
If the analysis algorithm finds an error
in the abstract state space,
then it stops and returns the yet incomplete
sets $\reached$ and $\wait$.
Now the corresponding abstract error path is extracted 
from the set $\reached$,
using the procedure $\mathsf{ExtractErrorPath}$,
and passed to the procedure $\mathsf{IsFeasible}$
for the \emph{feasibility check}.
If the abstract error path is feasible,
meaning there exists a corresponding concrete error path, 
then this error path represents a violation of the specification
and the algorithm terminates,
reporting \unsafe. 
If the error path is infeasible,
i.e., is not corresponding to a concrete program path,
then the precision was too coarse
and needs to be refined.
The \emph{refinement} step is performed
by the procedure $\mathsf{Refine}$ (cf.~Alg.~\ref{algorithm:Refine})
which returns a precision $\pr$
that makes the analysis strong enough
to exclude the infeasible error path
from future state-space explorations.
This returned precision is used to extend
the current precision of the \cegar algorithm,
which starts its next iteration,
delegating to the analysis algorithm
the re-computation of the sets $\reached$ and $\wait$
based on this refined precision.
%
\cegar is often used with lazy abstraction~\cite{LazyAbstraction}
to avoid re-discovering the whole state space after each refinement,
but instead removing only those parts of $\reached$ and $\wait$
that need to be re-analyzed with the new precision.

\subsection{Interpolation for Constraint Sequences}
\begin{algorithm}[t]
\vspace{1mm}
\begin{small}
\caption{$\mathsf{Refine}(\path)$}
\label{algorithm:Refine}
\begin{algorithmic} [1] 
\INPUT{an infeasible error path $\path = \seq{(\op_1, \pc_1), \ldots, (\op_n, \pc_n)}$}
\vspace{1mm}
\OUTPUT{a precision $\pr$}
\vspace{1mm}
\VARDECL{a constraint sequence $\cseqintpol$}

\vspace{1mm}
\STATE{$\cseqintpol := \seq{}$}
\STATE{$\pr(\pc) := \{\}$, for all program locations $\pc$}
\FOR{$i := 1$ to $n-1$}
  \STATE{$\cseq^+ := \seq{\op_{i+1}, \ldots, \op_n}$}
  \STATE{$\cseqintpol := \mathsf{Interpolate}(\cseqintpol \land \op_i, \cseq^+)$} \hspace{0mm} // inductive interpolation
  \STATE{$\pr(\pc_i) := \mathsf{ExtractPrecision}(\cseqintpol)$} \hspace{2mm} // create precision based on $\cseqintpol$
\ENDFOR
\STATE{\bf return $\pr$}
\end{algorithmic}
\end{small}
\vspace{1mm}
\end{algorithm}
An \emph{interpolant} for two constraint sequences
$\cseq^-$ and~$\cseq^+$, such that $\cseq^- \land \cseq^+$ is contradicting,
is a constraint sequence~$\cseqintpol$
for which
\begin{inparaenum}[1\upshape)]
	\item the implication $\cseq^- \implies \cseqintpol$ holds,
	\item the conjunction $\cseqintpol \land \cseq^+$ is contradicting, and
	\item the interpolant $\cseqintpol$ contains in its constraints only variables
	       that occur in both $\cseq^-$ and $\cseq^+$~\cite{CPAexplicit}.
\end{inparaenum}

In the following, we will introduce our novel approach,
which extends the procedure~$\mathsf{Refine}$
to not only perform interpolation
on a single infeasible error path,
and returning an arbitrary interpolant,
but instead, interpolate a set of infeasible sliced prefixes
stemming from this single infeasible error path,
and offering a set of interpolants
from which the most suitable precision may be chosen.

\addtolength{\textfloatsep}{-\baselineskip}

\section{Sliced Prefixes}
Our novel technique
can be used to extend any approach that is based on \cegar.
\emph{Slice-based refinement selection} extracts from a given infeasible error path
not only one single interpolation problem for obtaining a refined precision,
but a set of (more abstract, sliced) infeasible error paths
and thus a set of interpolation problems, from which the refined precision can be derived.
The interpolation problems for the extracted paths are given, one by one,
to the interpolation engine,
in order to derive interpolants for each path individually.
Hence, the abstraction refinement of the analysis is no longer dependent
on what the interpolation engine produces,
but instead it is free to choose from a set of interpolants
the one it finds most suitable.
The move from solving a single interpolation problem
to solving multiple interpolation problems to enable refinement selection,
and in the process transforming the refinement selection
into an optimization problem, is a key insight of our approach.

\subsection{Infeasible Sliced Prefixes}
A \cegar-based analysis usually encounters
an infeasible error path due to the coarse precision that it starts with.
%
This occurs when there exists a path to the error location
that contains as least one assume operation
that is feasible when the reachability algorithm computes \emph{abstract} successors
based on the current precision,
but is actually contradicting under the \emph{concrete} semantics of the program.
%
Every infeasible error path contains at least one such contradicting assume operation,
but often,
there exist several independent contradicting assume operations
in an infeasible error path, which leads to the notion of infeasible sliced prefixes:
A path~$\phi = \seq{(\op_1, \pc_1), \ldots, (\op_w, \pc_w)}$
is a \emph{sliced prefix} for a program path~$\path = \seq{(\op_1, \pc_1), \ldots, (\op_n, \pc_n)}$ if
$w \leq n$ and for all ${1 \leq i \leq w}$, we have 
  $\phi.\pc_i = \path.\pc_i$ $\mathrm{~and~}$ ${(\phi.\op_i = \path.\op_i \mathrm{~or~} 
     (\phi.\op_i = [true] \mathrm{~and~} \path.\op_i \mathrm{~is~assume~op}))}$,
i.e., a sliced prefix results from a path by
omitting pairs of operations and locations from the end, and possibly replacing some assume operations by
no-op operations.
If a sliced prefix for $\path$ is infeasible, then $\path$ is infeasible.

\subsection{Extracting Infeasible Sliced Prefixes from an Infeasible Error Path}
Algorithm~\ref{algorithm:ExtractSlicedPrefixes} is capable of extracting
from an infeasible error path
all its infeasible sliced prefixes,
i.e., all paths from the initial program operation to a contradicting assume operation.
%
The algorithm iterates through the given infeasible error path~$\path$.
It keeps incrementing a sliced path prefix~$\path_f$
that contains all operations from~$\path$ that were seen so far,
except the contradicting assume operations,
which are replaced by no-op operations.
Thus, $\path_f$~always stays feasible.
For every element~$(\op,\pc)$ from the original path~$\path$ 
(iterating in order from the first to the last pair),
we check whether it contradicts~$\path_f$,
which is the case if the result of the strongest-post operator
for the path $\path_f \land (\op,\pc)$ is contradicting (denoted by $\bot$).
If so, the algorithm has found a new infeasible sliced prefix.
In any case, it continues with the next element
after extending~$\sigma_f$
(either by the current operation, or by a no-op operation if the current operation is contradicting).
When the algorithm terminates,
which is guaranteed because $\path$~is finite,
the set~$\Sigma$ contains all infeasible sliced prefixes of~$\path$.
There is always at least one infeasible sliced prefix
because $\path$~is infeasible.
\begin{algorithm}[t]
\begin{small}
\caption{ExtractSlicedPrefixes($\path$)}
\label{algorithm:ExtractSlicedPrefixes}
\begin{algorithmic}[1] 
\INPUT{an infeasible path~$\path = \seq{(\op_1, \pc_1), \ldots, (\op_n, \pc_n)}$}
\vspace{1mm}
\OUTPUT{a non-empty set $\Sigma = \{\path_1, \ldots, \path_n\}$ of infeasible sliced prefixes of $\path$}
\vspace{1mm}
\VARDECL{a path $\path_f$ that is always feasible}
\vspace{1mm}
\STATE{$\Sigma := \{\}$;
$\path_f := \seq{}$}
\FOR{{\bf each} $(\op, \pc) \in \sigma$  // iterate in order from $(\op_1, \pc_1)$ to $(\op_n, \pc_n)$}
  \IF{$\SP{\path_f \land (\op, \pc)}{\top} = \bot$}
    \STATE{// add $\path_f \land (\op, \pc)$ to the set of infeasible sliced prefixes}
    \STATE{$\Sigma := \Sigma \cup \{\path_f \land (\op, \pc)\}$}
    \STATE{$\path_f := \path_f \land ([true], \pc)$}  \hspace{0.5mm} // append no-op
  \ELSE
    \STATE{$\path_f := \path_f \land (\op, \pc)$}  \hspace{5mm} // append original pair
  \ENDIF
\ENDFOR
\STATE{\bf return $\Sigma$}
\end{algorithmic}
\end{small}
\vspace{1mm}
\end{algorithm}

Algorithm~\ref{algorithm:ExtractSlicedPrefixes} returns the set of all infeasible sliced prefixes.
Each of these sliced prefixes has some interesting characteristics:
(1)~Each sliced prefix~$\phi$ starts with the initial operation~$op_1$,
and ends with an assume operation
that contradicts the previous operations of the sliced prefix,
i.e., $\SP{\phi}{\top} = \bot$.
(2)~The $i$-th sliced prefix,
excluding its (final and only) contradicting assume operation and location,
is a prefix of the $(i + 1)$-st sliced prefix.
(3)~All sliced prefixes differ from a prefix
of the original infeasible error path~$\path$
only in their no-op operations.

\begin{figure*}
\vspace{-3mm}
\centering
\begin{subfigure}[t]{.261\textwidth}
  \centering
    \vspace{0pt} 
    \includegraphics[scale=0.51]{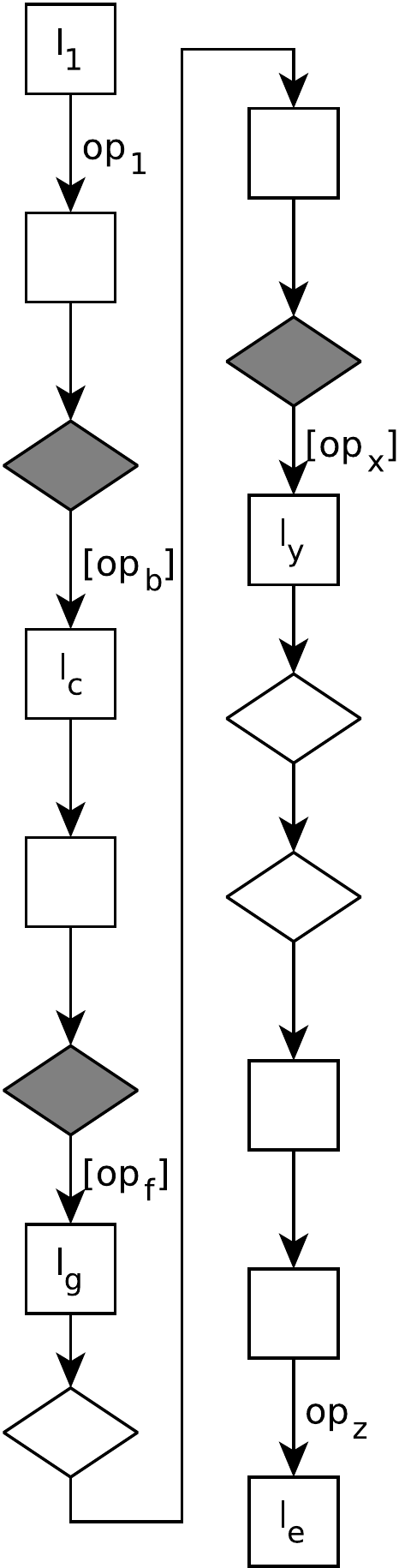}
  \caption{error path}
  \label{figure:spuriousCounterexample}
\end{subfigure}%
\hfil
\begin{subfigure}[t]{.36\textwidth}
  \centering
    \vspace{0pt}
    \includegraphics[scale=0.51]{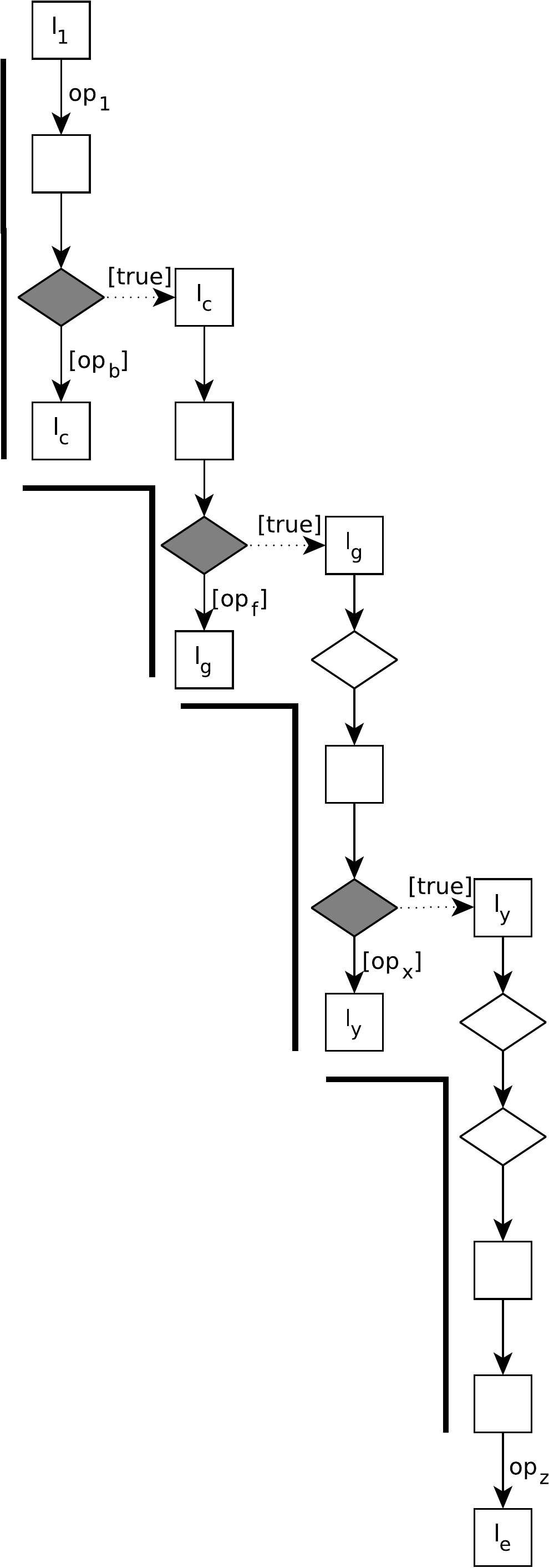}
  \caption{cascade of sliced prefixes}
  \label{figure:cascadeOfPrefixes}
\end{subfigure}%
\hfil
\begin{subfigure}[t]{.252\textwidth}
  \centering
    \vspace{0pt}
    \includegraphics[scale=0.51]{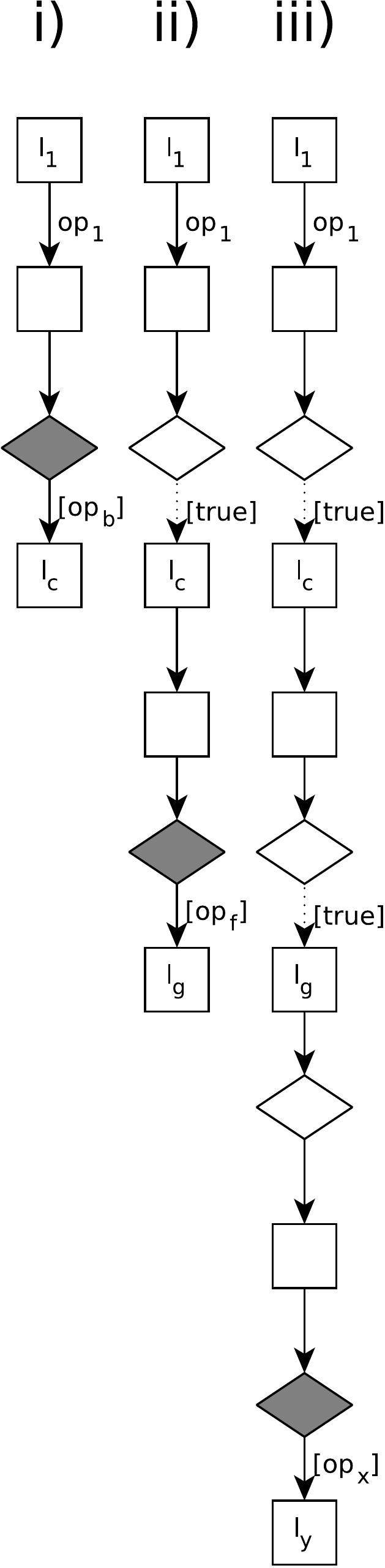}
  \caption{sliced prefixes}
  \label{figure:setOfPrefixes}
\end{subfigure}
\vspace{-1mm}
\caption{From one infeasible error path to a set of infeasible sliced prefixes}
\label{figure:cascadingInterpolation}
\vspace{-5mm}
\end{figure*}

The visualizations in Fig.~\ref{figure:cascadingInterpolation}
capture the details of this process.
Figure~\ref{figure:spuriousCounterexample} shows the original error path.
Nodes represent program locations
and edges represent operations
between these locations
(assignments to variables
or assume operations over variables,
the latter denoted with brackets).
To allow easier distinction,
program locations that are followed by assume operations
are drawn as diamonds,
while other program locations are drawn as squares.
Contradicting assume operations are drawn with a filled background.
The sequence of operations ends in the error state,
denoted by $l_e$.
Figure~\ref{figure:cascadeOfPrefixes}
depicts the cascade-like sliced prefixes that
the algorithm encounters during its progress.
Figure~\ref{figure:setOfPrefixes} shows the three infeasible sliced prefixes
that Alg.~\ref{algorithm:ExtractSlicedPrefixes} returns for this example.

The refinement procedure can now use any of these sliced prefixes to create interpolation problems,
and is not bound to a single sequence of interpolants for a single infeasible error path;
a refinement selection from different precisions is now possible.
The following proposition states that this is a valid refinement process.

\enlargethispage{1mm}
\subsubsection{Proposition} 
Let $\path$ be an infeasible error path and $\phi$ be the $i$-th infeasible sliced prefix
for $\path$ that is extracted by Alg.~\ref{algorithm:ExtractSlicedPrefixes},
then all interpolant sequences for $\phi$ are also interpolant sequences for~$\path$.

\subsubsection{Proof}\label{proof:infeasiblePrefixes} 
Let $\path~=~\seq{(\op_1, \pc_1), \ldots, (\op_n, \pc_n)}$
and $\phi~=~\seq{(\op_1, \pc_1), \ldots, (\op_w, \pc_w)}$.
Let $\cseqintpol_{\phi^j}$
be the $j$-th interpolant of an interpolant sequence for~$\phi$, 
i.e., for the two constraint sequences $\cseq^{-}_{\phi^j}~=~\seq{\op_1, \ldots, \op_j}$
and $\cseq^{+}_{\phi^j}~=~\seq{\op_{j+1}, \ldots, \op_w}$,
with $1 \leq j < w$.
Because $\phi$ is infeasible,
the two constraint sequences $\cseq^{-}_{\phi^j}$
and $\cseq^{+}_{\phi^j}$ are contradicting,
and therefore, $\cseqintpol_{\phi^j}$ exists~\cite{CPAexplicit}.
The interpolant $\cseqintpol_{\phi^j}$ is also an interpolant for
$\cseq^{-}_{\path^j} = \seq{\op_1, \ldots, \op_j}$
and $\cseq^{+}_{\path^j} = \seq{\op_{j+1}, \ldots, \op_e}$,
if
\begin{inparaenum}[\upshape(1\upshape)]
  \item \label{item:proofItp1} the implication $\cseq^{-}_{\path^j}~\implies~\cseqintpol_{\phi^j}$ holds,
	\item \label{item:proofItp2} the conjunction $\cseqintpol_{\phi^j}~\land~\cseq^{+}_{\path^j}$ is contradicting, and
	\item \label{item:proofItp3} the interpolant $\cseqintpol_{\phi^j}$ contains only variables
	       that occur in both $\cseq^{-}_{\path^j}$ and $\cseq^{+}_{\path^j}$.
\end{inparaenum}
Consider that $\cseq^{-}_{\phi^j}$ was created from $\cseq^{-}_{\path^j}$
by replacing some assume operations by no-op operations,
and that $\cseq^{+}_{\phi^j}$ was created from $\cseq^{+}_{\path^j}$
by replacing some assume operations by no-op operations
and by removing the operations $\seq{\op_{w+1},\ldots,\op_{n}}$ at the end.
Thus, both $\cseq^{-}_{\phi^j}$ and $\cseq^{+}_{\phi^j}$
do not contain any additional constraints (except for no-op operations)
than $\cseq^{-}_{\path^j}$ and $\cseq^{+}_{\path^j}$, respectively.
%
Because $\cseqintpol_{\phi^j}$ is an interpolant for $\cseq^{-}_{\phi^j}$ and $\cseq^{+}_{\phi^j}$,
we know that $\cseq^{-}_{\phi^j} \implies \cseqintpol_{\phi^j}$ holds,
and because $\cseq^{-}_{\path^j}$ can only be stronger than $\cseq^{-}_{\phi^j}$,
Claim~(\ref{item:proofItp1}) follows.
The conjunction $\cseqintpol_{\phi^j} \land \cseq^{+}_{\phi^j}$ is contradicting,
and $\cseq^{+}_{\path^j}$ can only be stronger than $\cseq^{+}_{\phi^j}$.
Thus, Claim~(\ref{item:proofItp2}) holds.
Because $\cseqintpol_{\phi^j}$ references only variables
that occur in both $\cseq^{-}_{\phi^j}$ and $\cseq^{+}_{\phi^j}$,
which do not contain more variables than
$\cseq^{-}_{\path^j}$ and $\cseq^{+}_{\path^j}$, resp.,
Claim~(\ref{item:proofItp3}) holds.

\begin{algorithm}[t!]
\vspace{1mm}
\begin{small}
\caption{$\mathsf{Refine^+}(\path)$}
\label{algorithm:RefinePlus}
\begin{algorithmic} [1] 
\INPUT{an infeasible error path $\path = \seq{(\op_1, \pc_1), \ldots, (\op_n, \pc_n)}$}
\vspace{1mm}
\OUTPUT{a precision $\pr$}
\vspace{1mm}
\VARDECL{a constraint sequence $\cseqintpol$,\\
      a set $\Sigma$ of infeasible sliced prefixes of $\path$,\\
      a mapping $\iota$ from infeasible sliced prefixes and program locations to interpolants}

\vspace{1mm}
\STATE{$\Sigma := \mathsf{ExtractSlicedPrefixes}(\path)$}
\vspace{0.5mm}
\STATE{// compute interpolants for each location in each prefix}
\FOR{{\bf each} $\phi_j = \seq{(\op_1, \pc_1), \ldots, (\op_w, \pc_w)} \in \Sigma$}
  \STATE{$\cseqintpol := \seq{}$}
  \FOR{$i := 1$ to $w-1$}
    \STATE{$\cseq^+ := \seq{\op_{i+1}, \ldots, \op_w}$}
    \STATE{$\cseqintpol := \mathsf{Interpolate}(\cseqintpol \land \op_i, \cseq^+)$} \hspace{1.5mm} // inductive interpolation
    \STATE{$\iota(\phi_j, \pc_i) := \cseqintpol$}
  \ENDFOR
\ENDFOR
\vspace{1mm}
\STATE{// choose suitable sliced prefix}
\STATE{//   (based on the sliced prefixes and its interpolants)}
\STATE{$\phi_{selected} := \mathsf{ChooseSlicedPrefix}(\iota)$}
\vspace{1mm}
\STATE{// create precision based on chosen interpolants}
\STATE{$\pr(\pc) := \{\}$, for all program locations $\pc$}
\FOR{{\bf each} $(\op, \pc) \in \phi_{selected}$}
  \STATE{$\cseqintpol := \iota(\phi_{selected}, \pc)$}
  \STATE{$\pr(\pc) := \mathsf{ExtractPrecision}(\cseqintpol)$} \hspace{2mm} // create precision based on $\cseqintpol$
\ENDFOR

\STATE{\bf return $\pr$}
\end{algorithmic}
\end{small}
\vspace{1mm}
\end{algorithm}

\section{Slice-Based Refinement Selection}

As described earlier,
extracting good precisions from the infeasible error paths
is key to the \cegar technique,
and the choice of interpolants influences the quality of the precision, and thus, the effectiveness
of the analysis algorithm.
By using the results introduced in the previous section,
the refinement procedure can now be improved by selecting a precision that is
derived via interpolation from a selected sliced prefix.

Algorithm~\ref{algorithm:RefinePlus} shows
our algorithm for slice-based refinement selection,
which can be used as a replacement for Alg.~\ref{algorithm:Refine}
in the \cegar algorithm
and chooses a suitable interpolant sequence during the refinement step.
First, this algorithm uses $\mathsf{ExtractSlicedPrefixes}$
to extract all infeasible sliced prefixes.
Second, it computes interpolant sequences for all of them
and stores them in the mapping~$\iota$.
Third, one sliced prefix is chosen by a heuristic
(in function $\mathsf{ChooseSlicedPrefix}$)
and fourth, the returned precision is created from the interpolants
for the chosen sliced prefix.
The heuristic can decide based on the information
contained in the sliced prefixes
as well as in the interpolants,
e.g., which variables are referenced by the interpolants.

\subsection{Refinement-Selection Heuristics}
We regard the selection of interpolants for refinement
as an independent direction for further research,
but present several ideas on how to select interpolants here.
%
There are two obvious options for interpolant selection
that do not depend on the actual interpolants.
Using the interpolant sequence derived from the very first,
i.e., the shortest, infeasible prefix
may rule out many similar infeasible error paths.
The downside of this choice is
that the analysis has to track information very early,
possibly blowing up the state-space
and making the analysis less efficient.
%
The other straight-forward option
(also known as counterexample minimization~\cite{LazyAbstractionForArrays})
is to use the longest infeasible sliced prefix
(containing the last contradicting assume operation)
for computing an interpolant sequence.
This may lead to a precision
that is local to the error location
and does not require refining large parts of the state space
at the beginning of the error path.
However, it may also lead to a larger number of refinements
if many error paths with a common prefix exist.
%
A more advanced strategy is to analyze the domain types~\cite{DomainTypes}
of the variables that are referenced in the interpolant sequence.
Each interpolant sequence can be assigned a score
that depends on the domain types of the variables in the interpolant sequence
such that
the score of the interpolant sequence is better
if it references only `easy' types of variables, e.g., boolean variables,
and no integer variables or even loop counters.
This allows to focus on variables that are inexpensive to track,
avoid loop unrolling where possible,
and keep the size of the abstract state space as small as possible.
Furthermore, it is possible to estimate,
by means of the use-def relation of the variables in the interpolants,
how much of the already explored state-space has to be recomputed
depending on which interpolant sequence is chosen.
Based on that insight,
we can identify the interpolant sequence
that would ensure that only as little as possible
from the state space needs to be re-explored.
In addition to that,
many different refinement heuristics are conceivable.
For example, it would be possible to avoid sliced error
paths that contain non-linear arithmetic
if using predicate abstraction with an SMT solver for linear arithmetic.

In general, any such heuristic can be used
without changing the overall algorithm,
but only the function~$\mathsf{ChooseSlicedPrefix}$
in Alg.~\ref{algorithm:RefinePlus}
needs to be replaced accordingly.
Using a selection heuristic specifically developed for programs encoding an event-condition-action system
improved the effectiveness of our tool \cpachecker
in the RERS challenge 2014
and allowed it to obtain two gold and one bronze medals,
as well as two special achievements%
\,\footnote{Results are available at \url{http://www.rers-challenge.org/2014Isola/}}.
This shows that optimizing the \cegar loop
by using domain knowledge in the refinement step
can be rewarding,
and that our approach provides a possibility to do so easily.
In the following, we present detailed results
for the effectiveness of our approach for a value analysis
with the heuristic based on domain types.

\section{Experiments}
We implemented our approach
in the open-source verification framework \cpachecker,
which is available online%
\,\footnote{\url{http://cpachecker.sosy-lab.org}}
under the Apache~2.0 license.
\cpachecker already has several analyses implemented
that can be used for program analysis
with \cegar and lazy abstraction.
We only extended the refinement process to work according to Alg.~\ref{algorithm:RefinePlus} ($\mathsf{Refine^+}$),
and did neither change the abstract domains nor the interpolation engines.
Our implementation is available in the source-code repository of \cpachecker.
The tool, the benchmark programs, the configuration files, and the complete results
are available on the supplementary web page%
\,\footnote{\url{http://www.sosy-lab.org/~dbeyer/cpa-ref-sel/}}.

\subsection{Setup}
We used the same experimental setup as in the
International Competition on Software Verification (SV-COMP'14)~\cite{SVCOMP14}:
machines with Intel Core i7-2600 quad-core CPUs with 3.4\,GHz,
a memory limit of 15\,GB, and a time limit of 15~min.
We limited each verification run to one CPU core,
because we are interested in the consumed CPU time and the consumed wall time was not important.

\subsection{Benchmark Programs}
For benchmarking we used the C~programs of the category ``DeviceDrivers64'' of SV-COMP'14.
This category contains 1\,428~large programs based on real-world Linux-kernel device drivers
with an average of 6\,045~lines of code per program.
We consider this category to be especially interesting
because our approach focuses on improving refinements in large programs
(with long and complex error paths, and many contradicting assume operations per error path).
Verification of device drivers is
a challenging research topic~\cite{LDV12}
and an important application domain~\cite{SLAM,LDV}.
For completeness, we also report the results for the 2\,626~programs of all categories of SV-COMP'14
except ``Concurrency'', ``HeapManipulation'', ``MemorySafety'', and ``Recursive'',
which rely on features that were not supported by the used configurations of our tool.

\subsection{Configurations}
Out of the several abstract domains that are supported by \cpachecker,
we choose the value analysis with refinement and lazy abstraction~\cite{CPAexplicit} for our experiments.
This abstract domain tracks explicit values for each program variable,
and in case the safety of the program depends on facts
that cannot be handled by the value analysis,
it delegates to an auxiliary predicate analysis,
which is configured for single-block-encoding~\cite{ABE}.
We used \cpachecker in revision~15\,509 of tag \texttt{cpachecker-1.3.10-refinementSelection}.

When using slice-based refinement selection,
the heuristic for choosing sliced prefixes
(function $\mathsf{ChooseSlicedPrefix}$ in Alg.~\ref{algorithm:RefinePlus})
was configured to select the interpolant sequence with the best score
based on the domain types of the variables~\cite{DomainTypes} referenced in the interpolants,
i.e., variables with a boolean character are favored over integer variables and loop counters.

\begin{table}[t]
  \centering
  \normalsize
  \caption{Results for slice-based refinement selection}
  \begin{tabular}{l | r r | r r}%
    Tasks & \multicolumn{2}{c|}{DeviceDrivers64} & \multicolumn{2}{c}{All} \\
    ~  & \multicolumn{2}{c|}{(1\,428 tasks)} & \multicolumn{2}{c}{(2\,626 tasks)} \\ \cmidrule{2-3} \cmidrule{4-5}
    Configuration & Classic  & Sliced        & Classic  & Sliced        \\ \midrule
    \# Solved     & 1328     & \textbf{1375} & 1932     & \textbf{1996} \\
    CPU time (h)  & 28.4     & \textbf{16.9} & 171      & \textbf{156}  \\
  \end{tabular}
\label{table:explicitResults}
\end{table}

\subsection{Results}
We now compare the results
of running the analysis with both
a classic refinement algorithm (as in Alg.~\ref{algorithm:Refine})
and our new refinement algorithm that is based on sliced prefixes (using Alg.~\ref{algorithm:RefinePlus}).
Table~\ref{table:explicitResults} shows a summary of the results.
The new approach proves to be effective,
by solving a total of 1\,375 of 1\,428 programs correctly
in the category ``DeviceDrivers64''.
Compared to the existing approach,
it solves 47~more programs correctly
and verifies all programs that could be verified before, too
(no regressions).
At the same time, the total CPU time
was reduced to 60\,\%. 
The reason for this vast improvement
is that the heuristic for choosing sliced prefixes
(guided by the domain type of the referenced variables)
is especially effective for the highly complex
and heterogeneous program code
in Linux-kernel device drivers.
On the set of all programs,
slice-based refinement selection is effective, too.
It can solve 64~more verification problems correctly
and needs almost 10\,\% less time. 

\begin{figure}[t]
\vspace{-4mm}
\centering
\begin{subfigure}{0.49\linewidth}
  \includegraphics[width=\linewidth]{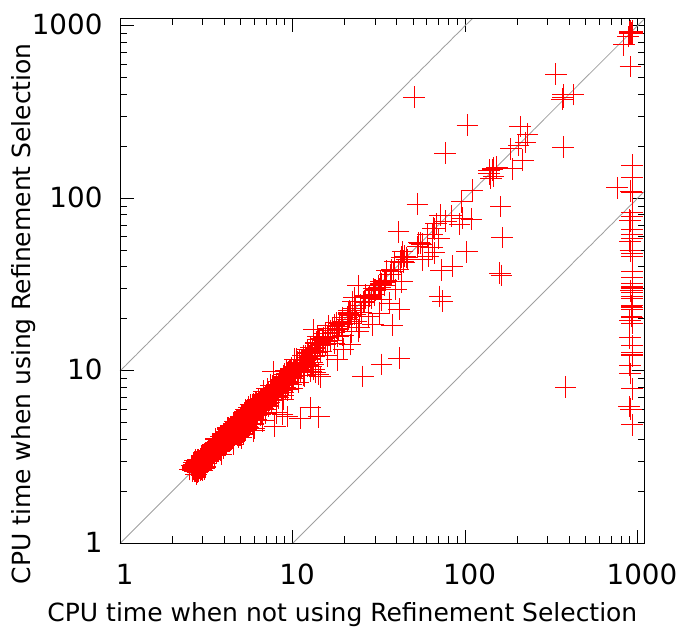}
  \caption{on category ``DeviceDrivers64''}
  \label{fig:explicitResults-dd64}
\end{subfigure}
\begin{subfigure}{0.49\linewidth}
  \includegraphics[width=\linewidth]{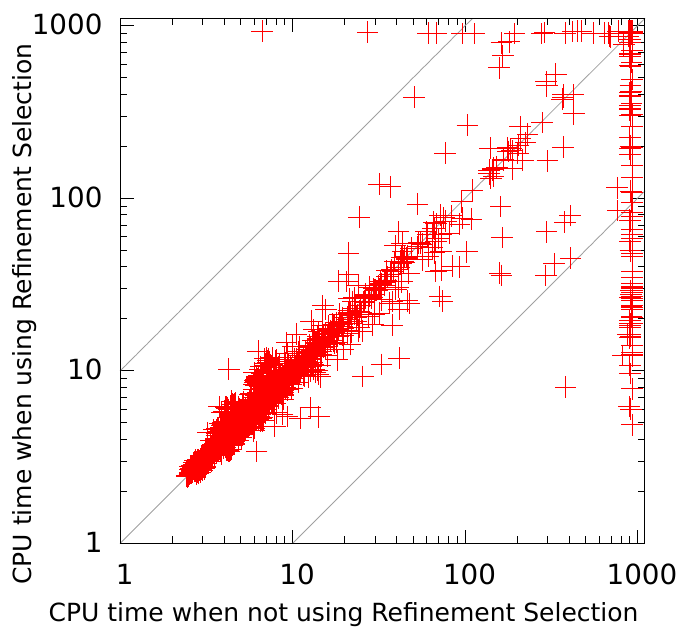}
  \caption{on all 2\,626 verification tasks}
  \label{fig:explicitResults-all}
\end{subfigure}
\caption{Scatter plots comparing the CPU when not using slice-based refinement selection (x-axis)
  with the CPU time when using slice-based refinement selection (y-axis)}
\label{fig:explicitResults}
\vspace{-3mm}
\end{figure}

Figure~\ref{fig:explicitResults} shows scatter plots
for comparing the CPU time of slice-based refinement selection
versus the existing approach on both sets of verification tasks.
Only data points for successful verification runs
and timeouts are shown (out-of-memory runs are omitted).
The figures show that our approach in many cases
makes the difference between solving the verification task
within the time limit, and not solving the verification task at all
(such instances are those at the right border of the plot).
This illustrates that without slice-based refinement selection
and our heuristic for avoiding loop counters in the precision,
the interpolants will sometimes be such that the analysis
has to unroll long loops, which causes state-space explosion;
this can often be avoided with the new approach.
The plot also shows that for most of the remaining programs
there is no difference in time.
This is due to the fact that both sets
also contain a large number of small programs,
for which our approach does not make a difference,
because the counterexamples are short and simple.
Figure~\ref{fig:explicitResults-dd64} shows that
for the category ``DeviceDrivers64'', there is not a single effectiveness regression,
i.e., all verification tasks that the classic approach can solve
can also be solved by slice-based refinement selection --- plus 47 more.
Figure~\ref{fig:explicitResults-all} shows that on the set of all programs,
there are a few regressions where there is a timeout
when using the new approach.
These are randomly created programs that belong to the ``ECA'' subset of SV-COMP'14.
All variables in these programs have the same domain type,
and thus, our heuristic for choosing interpolants based on the domain types of variables
is not effective here.
For this subset, a heuristic specifically developed for the ECA programs of RERS'14 
was successful.

\section{Conclusion}
In this work we presented our novel approach of \emph{slice-based refinement selection},
which extracts several infeasible sliced prefixes
from one single infeasible error path.
From any of these infeasible sliced prefixes,
an independent interpolation problem can be derived
that can be solved by a standard interpolation engine,
and the analysis can choose from the resulting interpolant sequences
the one thought to be best for the verification.
Our novel approach is independent from the abstract domain
(in particular, does not depend on an SMT solver)
and can be combined with any analysis that is based on 
\cegar and interpolation-based abstraction refinement, while
previous work on guided interpolation~\cite{ExploringInterpolants}
is applicable only to SMT-based approaches. 
We experimentally demonstrated that the novel approach
using a heuristic based on domain types can significantly
improve the effectiveness and efficiency of the program analysis.
We also discussed some possible further heuristics to select suitable interpolant sequences.

\balance

\bibliography{dbeyer,sw,tah}

\begin{acronym}
  \acro{CEGAR}{Counterexample-Guided Abstraction Refinement}
  \acro{CFA}{control-flow automaton}
  \acro{ARG}{abstract reachability graph}  
  \acro{SMT}{Satisfiability Modulo Theories}
  \acro{ABE}{Adjustable-Block Encoding}
\end{acronym}

\end{document}